\documentclass[useAMS,usenatbib]{mn2e}
\usepackage{times}
\usepackage{graphicx}
\title[Large-scale clustering of faint red galaxies]
{Understanding the faint red galaxy population using large-scale clustering measurements from SDSS DR7}

\author[A. J. Ross, R. Tojeiro, \& W. J. Percival]{Ashley J. Ross\thanks{Email: Ashley.Ross@port.ac.uk}, Rita Tojeiro, \& Will J. Percival \\
Institute of Cosmology \& Gravitation, Dennis Sciama Building, University of Portsmouth, Portsmouth, PO1 3FX, UK\\}

\begin{document}

\date{Accepted for Publication by MNRAS}

\pagerange{\pageref{firstpage}--\pageref{lastpage}} \pubyear{2010}

\maketitle

\label{firstpage}

\begin{abstract}

We use data from the SDSS to investigate the evolution of the large-scale galaxy bias as a function of luminosity for red galaxies. We carefully consider correlation functions of galaxies selected from both photometric and spectroscopic data, and cross-correlations between them, to obtain multiple measurements of the large-scale bias. We find, for our most robust analyses, a strong increase in bias with luminosity for the most luminous galaxies, an intermediate regime where bias does not evolve strongly over a range of two magnitudes in galaxy luminosity, and no evidence for an upturn in bias for fainter red galaxies. Previous work has found an increase in bias to low luminosities that has been widely interpreted as being caused by a strong preference for red dwarf galaxies to be satellites in the most massive halos. We can recover such an upturn in bias to faint luminosities if we push our measurements to small scales, and include galaxy clustering measurements along the line-of-sight, where we expect non-linear effects to be the strongest. The results that we expect to be most robust suggest that the low luminosity population of red galaxies is not dominated by satellite galaxies occupying the most massive haloes.

\end{abstract}


\section{Introduction}\label{sec:intro}

The evolution of dark matter can be accurately modelled within the $\Lambda$ Cold Dark Matter ($\Lambda$CDM) scenario, giving the number density of collapsed objects (e.g., dark matter halos) of given mass as a function of redshift. In the widely accepted model of galaxy evolution, galaxies form within the gravitational potential wells of host dark matter haloes \citep{WhiteRees1978}. Thus, we can reduce the problem of modelling galaxy populations to one of working out the set of assumptions on how to relate luminous with dark matter. In fact, simple instructions can provide an excellent description of the observed clustering of galaxies in the near (see, e.g., \citealt{N02,Z10}) and in the more distant universe (see, e.g., \citealt{Coil06,Mc07}).

In the galaxy formation framework, structure builds hierarchically, with smaller halos being the first to collapse and merge together to build up larger haloes over time. Galaxies, as permanent residents of these haloes, must therefore also grow through merging, but the link between galaxy growth (in terms of stellar mass) and halo growth (in terms of dark matter mass) is far from direct. Firstly, stellar mass in galaxies may also grow through the process of turning cold gas into stars, and secondly galaxy growth through merging is slower than halo growth - galaxies do not always merge when their host haloes do, and there is strong evidence that stellar mass may not be conserved during such a process (\citealt{MW07, ConroyEtAl07}). 

There is a firmly-established colour bimodality in the galaxy population seen at low redshift (e.g. \citealt{BlantonEtAl03}), and this continues to earlier epochs (e.g, \citealt{Bell04}). This striking separation of the galaxy population into a blue (star-forming) and red (quiescent) clouds must therefore be an outcome of any galaxy evolution model. Understanding how these two populations evolve, and how galaxies go from one cloud to the other (i.e. what process quenches star formation), has been a long-standing quest in galaxy evolution, and one for which clustering studies have the potential of being particularly insightful.

The large-scale clustering strength of a particular population of galaxies is directly linked to the host dark matter halo mass \citep{BBKS,cole89}. Consequently, studying the clustering properties of a sample of galaxies, as well as its evolution with redshift, remains as one of the most powerful ways to constrain galaxy evolution. Many previous studies have investigated the clustering of galaxies as a function of their colour and luminosity (recent studies include, e.g., \citealt{W98,N02,Ma03,Z05,Cr06,R07,M08,R09,T10,Z10}). In broad terms, it has become clear that the large-scale clustering strength increases with luminosity, and with redder colour.  

Clustering measurements of galaxy sub-populations, in one sense, represent statistical restatements and refinements of the well known morphology-density relationship \citep{Dre80}, which is now known more accurately as a colour-density relationship (see, e.g., \citealt{Ball08,Skibba08}). We now know that the bi-modality in the colours of galaxies extends to their large-scale clustering strengths, revealing that a halo-mass dependent effect is important in delineating blue from red galaxies. Essentially, the most luminous, most red galaxies occupy the highest peaks in the density fields, and for any particular luminosity, red galaxies are found to occupy higher peaks than their blue counterparts.  This suggests red galaxies were the first to begin their hierarchical mass assembly --- qualitatively consistent with the interpretation that they are red because they have used up/been stripped of their cold gas (see, e.g., \citealt{Larson80,Cow96}). 

If we now focus on the clustering of red galaxies, there is interesting evidence that their large-scale clustering strength increases towards the faint end (see, e.g., \citealt{N02,Z05,swanson, cresswell}). The natural interpretation is that faint red galaxies are predominantly satellite galaxies in large mass haloes. This would further imply that red galaxies with luminosities $\sim$$L^{*}$ prefer less massive halos than their brighter and fainter counterparts --- qualitatively consistent with the picture of S0 galaxies occupying groups and luminous and dwarf red galaxies occupying clusters.

While it is uncontroversial that a significant percentage of faint red galaxies are satellites in large mass haloes, the exact proportion that are satellites, and whether this is sufficient that these galaxies dominate the population, is less clear.  \cite{Wang09}, selecting on colour, find that dwarf red galaxies (fainter than those studied in any of the other cited works here) are predominantly central, rather than satellite galaxies.  This is in contrast to \cite{Haines}, who, when selecting on H$\alpha$ emission, find nearly all red dwarf galaxies are satellites.  Furthermore, there is some tension between analyses of large-scale bias and the work of \cite{Mas08}, as they find that faint red galaxies have a very weak contribution to this growth of luminous red galaxies (LRGs), i.e. if the faint galaxies occupy the same halos as LRGs, they are somehow precluded from merging with them. Finally, \cite{Brown} do not see an upturn in the large-scale bias towards fainter luminosities when studying red galaxies in the Bootes field, using photometric redshifts.

Motivated partly by the tension described in the previous paragraph, and partly by the desire to study fainter red galaxies, in this work we take a new look at the large-scale bias of red galaxies in the SDSS seventh and final data release, using both photometric and spectroscopic data. The photometric data allows us to analyse a significantly larger volume of the Universe containing the lower luminosity galaxies than the spectroscopic data. This means we can consider the same volume to analyse a wide range of galaxy luminosities. We also use the cross-correlation between these data with brighter spectroscopic data to confirm the robustness of our results.

This paper is organised as follows:  In Section~\ref{sec:data} we describe the photometric and spectroscopic redshift catalogs we use for our measurements.  In Section~\ref{sec:meth} we describe how we measure correlation functions (angular auto- and cross-correlation functions and redshift space 3D auto-correlation functions) and how we use these measurements to measure the bias of galaxy samples.  In Section~\ref{sec:photb}, we present the bias measurements from our photometric redshift samples, and in Section~\ref{sec:bspec} from our spectroscopic redshift samples.  In Section~\ref{sec:com}, we compare our spectroscopic and photometric results to each other and to the results of previous studies.  In Section~\ref{sec:phys}, we discuss the physical implications of our measurements and in Section~\ref{sec:con}, we present a summary of our conclusions.

Throughout, we use $M_r$ as shorthand for $M_r - 5{\rm log}(h)$. We assume a flat cosmology with $\Omega_m = 0.25$, $h = 0.7$, $\sigma_8 = 0.8$, and $\Omega_b = 0.045$.

\section{Data} \label{sec:data}

We use data from the SDSS seventh data release (DR7).  This survey obtained wide-field CCD photometry (\citealt{C}) in five passbands ($u,g,r,i,z$; e.g., \citealt{F}), amassing nearly 10,000 square degrees of imaging data for which object detection is reliable to $r \sim 22$ (\citealt{DR7}).  From these photometric data, spectroscopic targets have been identified and observed yielding a sample of galaxies with over 600,000 spectroscopic redshifts complete to a \cite{Petrosian} magnitude limit of $r < 17.77$ occupying over 8,000 square degrees (\citealt{DR7}).  We utilize data from both the spectroscopic redshift catalog (as this allows precise redshift information) and a photometric redshift catalog (as this allows us to go more than 2 magnitudes deeper).

\begin{figure}
\includegraphics[width=0.9\columnwidth]{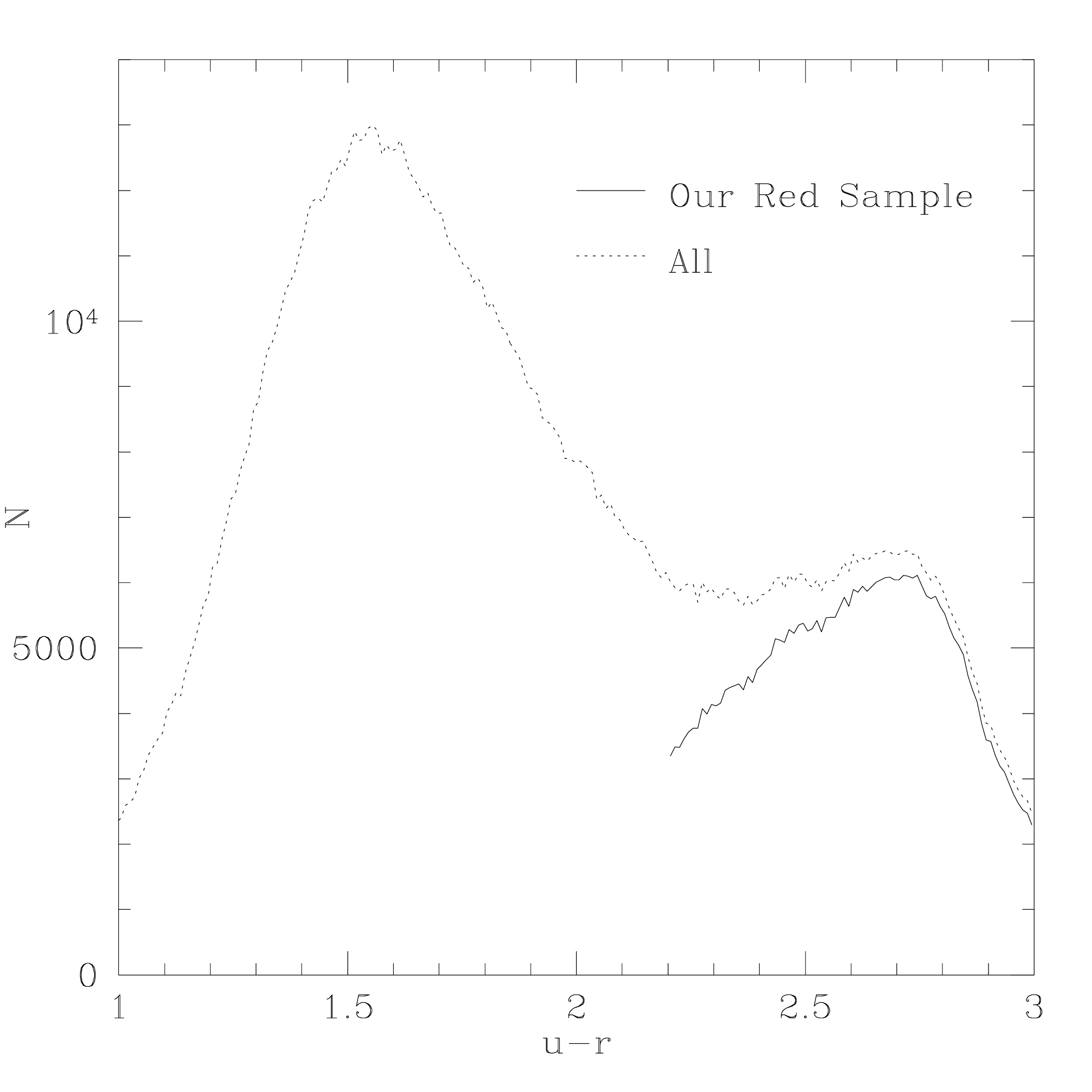}
\caption{The distribution of $u-r$ colours of galaxies with $z_{phot} < 0.1$ and $M_r < -17.75$ (dotted line) and for galaxies with only $u-r > 2.2$ and $g-r > 0.8$ (solid line). }
\label{fig:ccut}
\end{figure}

The photometric redshifts used were obtained from the SDSS {\rm photoz} table. Redshifts in this catalog were estimated using a hybrid template/empirical approach, and the output includes rest-frame absolute magnitudes, k-corrections, and galaxy-type values. We include all galaxies with $z_{phot} < 0.1$ and $M_r < -17.75$ (which makes our sample approximately volume limited for $r < 20$).  In order to select red galaxies, we use the (k-corrected) double colour cut
\begin{equation}
u-r > 2.2~,~ g-r > 0.8,
\label{eq:cc}
\end{equation}
which ensures that the galaxies are truly red and there are few blue interlopers in our sample. Fig.~\ref{fig:ccut} displays the distribution of $u-r$ colours of the galaxies in our red sample, with a solid line.  The dashed displays the distribution we would have had if we simply selected $u-r > 2.2$.  Our double cut clearly removes galaxies that are close to our cut limit, giving us a distinct sample of galaxies (more so than selecting on type value provided in the {\rm photoz} table).  We find that performing this double-cut also improves the reliability of the photometric redshift estimates.  We further cut the data to the angular mask of Ross et al. (2010; hereafter R10), leaving just over 6000 square degrees of observing area in the Northern contiguous area of SDSS, and subdivide this sample into three subsamples with $-17.75 < M_r < -18.75$ (27023 galaxies), $-18.75 < M_r < -19.75$ (23081 galaxies), and $-19.75 < M_r < -20.75$ (17156 galaxies).

We also consider galaxies selected from the main SDSS spectroscopic sample. The red galaxy samples were selected using the colour cut as defined by Eq.~(\ref{eq:cc}), dividing into five subsamples, by absolute magnitude, $-17 < M_r < -18$ (2082 galaxies), $-18 < M_r < -19$ (11663 galaxies), $-19 < M_r < -20$ (54908 galaxies), $-20 < M_r < -21$ (118853 galaxies), and $-21 < M_r < -22$ (97947 galaxies).  Note that we place no redshift limits on the red galaxy samples we use. 

We also consider spectroscopic samples without colour selection to cross-correlate against the photometric redshift samples. These samples were selected to be approximately volume limited; one is limited to $z < 0.15, M_r < -20.75$ (129435 galaxies) and a second limited to $z < 0.1, M_r < -19.75$ (140341 galaxies).  

\section{Methodology} \label{sec:meth}

We now outline the methodology used to calculate the bias of red galaxies as a function of their absolute magnitude. For the photometric data and for cross-correlations with the spectroscopic data, we use the projected angular correlation function. For the spectroscopic data we consider the full 3D correlation function.

\subsection{Calculating Angular Correlation Functions}

In order to calculate angular auto-correlation and cross-correlation functions, $w(\theta)$, we use the \cite{LZ} estimator 
\begin{equation}
w(\theta) = \frac{D_1D_2(\theta) - D_1R_2(\theta)  - D_2R_1(\theta)+ R_1R_2(\theta)}{R_2R_2(\theta)},
\label{eq:LZ}
\end{equation}
where $D_1$ and $D_2$ (and $R_1$ and $R_2$) represent separate data samples for the cross-correlation and are the same for the auto-correlation.  Our random catalog representing the photometric data contains 10 million points and is the same as used in R10. The photometric catalogs (data and random) are masked for seeing, reddening, bright stars, satellite trails, etc. in the same manner as R10.  At scales larger than $1^{o}$, we employ a pixel based routine using SDSSpix.  We use the pixel resolution (64) such that the pixels have an area with equivalent circular radius of $0.083^{\rm o}$. We confirm that the point to point and pixelized methods agree at overlapping scales. The random catalog representing the spectroscopic data contains over 6 million points and was created as described in \citep{reidetal09}, with a pixel based mask based on {\sc HEALPIX} \citep{GorskiEtAl05}.

We use a jackknife method (e.g., \citealt{Scr02}), with inverse-variance weighting to estimate the errors and covariance matrix (e.g., \citealt{Mye07}) for $w(\theta)$.  The method is nearly identical to the method described in detail in Ross et al. (2007; hereafter R07) and applied to DR7 data in R10.  The jackknife method works by creating many subsamples of the entire data set, each with a small part of the total area removed.  R07 found that 20 jack-knife subsamplings are sufficient to create a stable covariance matrix, and we therefore use 20.  These 20 subsamples are created by extracting a contiguous grouping of 1/20th of the unmasked pixels in 20 separate areas.  Our covariance matrix, $C_{w}$, is thus given by
\begin{equation}
\begin{array}{ll}
C_{i,j, w }=  C_{w}(\theta_i,\theta_j) & ~\\
= \frac{19}{20} \sum_{k=1}^{20}[\omega_{full}(\theta_i) - \omega_{k}(\theta_i)][\omega_{full}(\theta_j) - \omega_{k}(\theta_j)], & ~
\end{array}
\label{eq:JK}
\end{equation} 
\noindent where $\omega_{k}(\theta)$ is the value for the correlation measurement omitting the $k$th subsample of data and $i$ and $j$ refer to the $i^{th}$ and $j^{th}$ angular bin. The jackknife errors are simply the square-root of diagonal elements of the covariance matrix.  

\subsection{Calculating 3D Correlation Functions}

In order to calculate the 3D auto-correlation function, $\xi(r,\mu)$, we again employ the \cite{LZ} estimator but in this case $DD$, $DR$, and $RR$ are now functions of $r$, the physical separation we calculate given our assumed cosmology, and also $\mu$, the {\rm cosine} of the angle between the radial direction and the alignment of a galaxy pair.  Further, we must assign the randoms radial positions, which we do by sampling a fine resolution cubic spline fit to the galaxy redshift distribution. We also weight each galaxy according to the number density at its radial position \citep{fkp}: this weight is optimal if the galaxies Poisson sample the density field.    

We analytically estimate the errors and covariance of the $\xi(r,\mu)$ measurements. Given optimal weights, $w(r) = 1/(1+\bar{n}(r)P(k))$, and the power spectrum, $P(k)$, its error can be estimated by \citep{tegfish}:
\begin{equation}
\sigma_p^2(k) = \frac{(2\pi)^3P(k)^2}{V_kV_{eff}(k)},
\label{eq:perr}
\end{equation}  
where $V_{eff}(k) = \int \left[ \frac{\bar{n}(r)P(k)}{1+\bar{n}(r)P(k)}\right]^2d^3r$ and $V_k = 4\pi k^2 \Delta k$.  Given that the correlation function is just the Fourier transform of the power-spectrum, we can estimate the covariance, $C_{\xi}$, in the measured $\xi$ between bins centred at $r_1$ and $r_2$ as
\begin{equation}
C_{\xi}(r_1,r_2) = \frac{1}{2\pi^2}\int \frac{ dkP(k)^2{\rm sin}(r_1k) {\rm sin}(r_2k)}{r_1r_2V_{eff}(k)}+\delta_{r_1,r_2}/n_p,
\end{equation}
where $\delta$ is the Kronecker delta, and $n_p$ is the number of pairs included in the bin. Note that although this procedure ignores correlations induced by the sample window function, it should be sufficiently accurate for the analysis attempted in this work, as we only wish to measure the broad amplitude of the clustering rather than the relative clustering on different scales: for such measurements of amplitude, correlations between data are unimportant.

\subsection{Theoretical Modelling}

We model the non-linear power-spectrum using the fitting formulae of \cite{smith}, which we use in combination with the transfer function of \cite{EH} to include the effects of baryons.  Given this power-spectrum, we determine the isotropic 3-dimensional real-space correlation function $\xi(r)$ via Fourier transform.  We model the redshift-space correlation function as \citep{hamilton92}
\begin{equation}
\xi^s(\mu,r) = \xi_o(r)P_o(\mu)+\xi_2(r)P_2(\mu) +  \xi_4(r)P_4(\mu),
\label{eq:ximus}
\end{equation}
where
\begin{eqnarray}
  \xi_0(r) &=& (b^2+\frac{2}{3}bf+\frac{1}{5}f^2)\xi(r), \\
  \xi_2(r) &=& (\frac{4}{3}bf+\frac{4}{7}f^2)[\xi(r)-\xi'(r)], \\
  \xi_4(r) &=& \frac{8}{35}f^2[\xi(r)+\frac{5}{2}\xi'(r)-\frac{7}{2}\xi''(r)],
\label{eq:xirs}
\end{eqnarray}
$P_\ell$ are the standard Legendre polynomials, and
\begin{eqnarray}
  \xi'\equiv3r^{-3}\int^r_0\xi(r')(r')^2dr' \\ 
  \xi''\equiv5r^{-5}\int^r_0\xi(r')(r')^4dr',
\end{eqnarray}
$b$ is the large-scale bias of the galaxy population being considered.  We will be comparing our models to measurements limited to different maximum $\mu$ values.  Thus, the model will become
\begin{equation}
\xi^s(\mu_{max},r) = \int_0^{\mu_{max}} \xi^s(\mu,r) d\mu/\mu_{max}
\end{equation}

In order to calculate model $w(\theta)$, we must project $\xi^s(\mu,r)$ over the radial distribution of galaxy pairs in a particular sample (or samples in the case of cross-correlations).
\begin{equation}  \label{eq:w2}
  w(\theta) = \int dz_1 \int dz_2 n_{i}(z_1)n_{j}(z_2)
    \xi^s\left[\mu,{\bf r}(\theta,z_1,z_2)\right],
\end{equation}
where $n_{i}$ is the normalized redshift distribution of sample $i$ (and $i = j$ for the auto-correlation).  The galaxy separation ${\bf r}$ is a function of the angular separation of the galaxies $\theta$ and their redshifts $z_1$ and $z_2$ (as is $\mu$).  

\subsection{Modelling Redshift Distributions}
\label{sec:nz}
The photometric redshifts we use have significant error associated with each redshift estimate.  This implies that estimation of the true redshift distribution of the photometric redshift data is not trivial.  In order to estimate the true redshift distribution for each of our photometric samples, we follow the procedure outlined in R10.  Given a redshift estimate and its error, we assume a Gaussian PDF.  We then convolve this PDF with the luminosity function and volume element/redshift relationship to obtain a final PDF, which we then sample 10 times.  In this work, we find that adding a bias to the photometric redshifts (simply a shift in the mean) of +0.005 for each sample allows for the best agreement between all of our clustering measurements (see Section~\ref{sec:photb}).

We can construct $M_r$ distributions in a similar manner.  Given the $M_r$ of the redshift estimate, we can calculate the value of $M_r$ at any point along the redshift PDF based on the change in the distance modulus, thus obtaining a PDF of $M_r$ for each galaxy.  This allows us to determine the expected mean $M_r$ for any particular photometric redshift sample.  When we calculate angular cross-correlations between photometric and spectroscopic samples, the redshift ranges of the samples are not matched.  Thus, the distribution of luminosities of the galaxies that contribute to the measured clustering signal will not necessarily match that of either sample. In order to find the expected mean $M_r$ of galaxies contributing to the cross-correlation, we can simply ignore the parts of the redshift PDF outside the spectroscopic redshift bound and thereby determine the effective mean $M_r$ of photometric galaxies between any chosen spectroscopic redshift bounds.

\begin{figure*}
\begin{minipage}{7in}
\centering
\includegraphics[width=180mm]{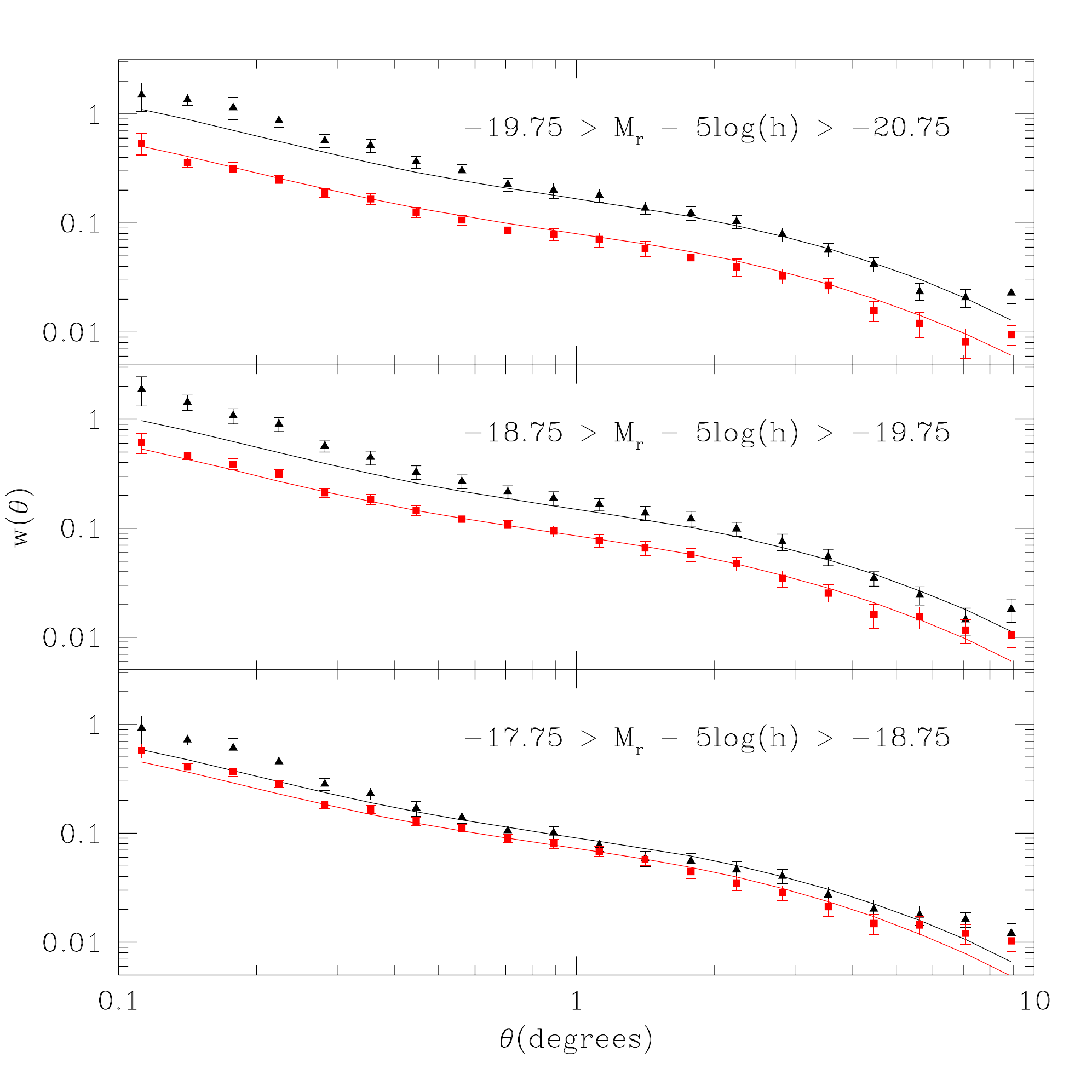}
\caption{The measured angular auto-correlations for the three specified photometric samples, all with $z_{phot} < 0.1$ (black triangles) and their cross-correlations (red squares) with the spectroscopic sample with $z < 0.15$, $M_r < -21.75$.  The black and red curves display the respective best-fit models (all fit between 1.0 and 5.0 degrees).}
\label{fig:photw}
\end{minipage}
\end{figure*}

\subsection{Calculating Bias Values}

We assume a simple linear bias model between the the 2-pt clustering in the dark matter and galaxy over-density fields, such that the bias $b$, is defined by
\begin{equation}
P(k)_g = b^2P(k)_{DM}.
\label{eq:bias}
\end{equation}
In the regime where linear theory is a good approximation this relationship naturally extends to $\xi(r)$ and $w(\theta)$.  The use of the \cite{smith} non-linear $P(k)$ has the potential to extend this relationship to weakly non-linear scales.  Thus, for arbitrary limits (but always assuming $b$ is scale independent in Eq. \ref{eq:bias}) we determine the best-fit value of $b$ for a particular galaxy sample by minimizing
\begin{equation}
\chi^2   =  \sum_{i,j}
  \left[w(\theta_i)-\hat{w}(\theta_i)\right]
  C_w^{-1}(\theta_i,\theta_j)
  \left[w(\theta_j)-\hat{w}(\theta_j)\right]
\end{equation}
where the model correlation function $\hat{w}$ and $\chi^2$ are implicitly dependent on the value of $b$.  For the 3D correlation functions, $w(\theta)$ is replaced by $\xi^s(r,\mu_{max})$ and $\theta$ by $r$.

\section{Bias from Photometric Data} \label{sec:photb}

We measure the bias of galaxies in our photometric samples by measuring their respective auto-correlations and cross-correlations with the spectroscopic sample of galaxies with $z_{spec} < 0.15$ and $M_r < -20.75$ (which we denote ``full"). The redshift limit means this sample covers most of the redshift distribution of the photometric samples and the absolute magnitude limit ensures that the bias of the galaxies in the sample should not change significantly with redshift.  Based on $w(\theta)$ measurements with $\theta > 1^{\rm o}$, the bias of the full spectroscopic sample is $1.32 \pm 0.04$, which agrees with the results of \cite{Z10}, as it lies between the bias values of their $M_r < -20.5$ and $M_r < -21$ samples.

We determine the mean $M_r$ of each photometric sample by estimating the distribution of $M_r$ values as described in section \ref{sec:nz}.  The cross-correlation with the spectroscopic sample implies a hard limit of $z < 0.15$ for the photometric galaxies contributing signal to the measured cross-correlations.  We find that imposing this limit when calculating the magnitude distribution changes the mean $M_r$ value by less than 0.05 magnitudes in every case. This implies that the $z < 0.15$ limit makes only a small difference to any of the magnitude distributions of the photometric galaxies that contribute to the cross-correlation signal. Therefore, it is valid to expect the bias of the photometric galaxies contributing to the auto-correlation signal to be the same as the bias of the photometric galaxies contributing signal to the cross-correlation with the $z < 0.15$ spectroscopic sample.  

Fig.~\ref{fig:photw} displays the measured $w(\theta)$ for our three main photometric samples.  The auto-correlations are shown by the black triangles and the cross-correlations are displayed in red-squares, with luminosity increasing from bottom to top.  The solid curves display the best-fit model $w(\theta)$ fit for $1^{\rm o} < \theta < 5^{\rm o}$, where we have jointly fit the $w(\theta)$ auto- and cross-correlations (note, bias values can be independently determined for the auto- and cross-correlations; the joint-fit determines the most consistent value and reduces the overall uncertainty).  We restrict the fits to $\theta < 5^{\rm o}$ because systematic effects in the data (due to, e.g., reddening and star/galaxy separation) and the modelling (effects of redshift space distortions) become more significant at larger angular scales.   The jack-knife errors increase only slightly over our range of measurements, due to competing effects of cosmic variance and shot-noise. The mean $M_r$ and best-fit $b$ values corresponding to these joint fits are -18.46, 1.2$\pm$0.05, -19.46, 1.43$\pm$0.07, and -20.39, 1.40$\pm$0.07, with respective $\chi^2$ values of 13.3, 16.0, and 17.0 (and 13 degrees of freedom).  We note that these values remain constant (within the 1$\sigma$ error-bars) regardless of the minimum scale we fit to between $0.5^{\rm o}$ and $2.0^{\rm o}$.  

In order to test the robustness of these results to the photometric redshift determination, we have also obtained estimates of $b$ for galaxies of different luminosity by cross-correlating the photometric samples with spectroscopic samples covering different redshift ranges. The cross-correlation signal depends on the number of galaxies in the photometric sample that lie within the redshift range covered by the spectroscopic sample, so testing whether these bias estimates match those from our full sample tests the assumed distribution of photometric redshifts. We cross-correlate each photometric sample with two additional spectroscopic samples; one with $z < 0.1$, $M_r < -19.75$ (which denote ``near") and second with $0.1 <  z < 0.15$, $M_r < -20.75$ (which we denote ``far"). We find that these redshift limits increase/decrease the mean value of $M_r$ of the galaxies contributing to the cross-correlation signal by $\sim 0.4$ magnitudes.  We find the bias of the near sample is 1.17 $\pm$ 0.05 (again in agreement with \citealt{Z10}) and we continue to use $b = 1.32 \pm 0.04$ for the far sample. Note that the near and far bias values can be determined only from the cross-correlations, since auto-correlations of the photometric data cannot replicate the effects of a hard redshift cut.

\begin{figure}
\includegraphics[width=0.9\columnwidth]{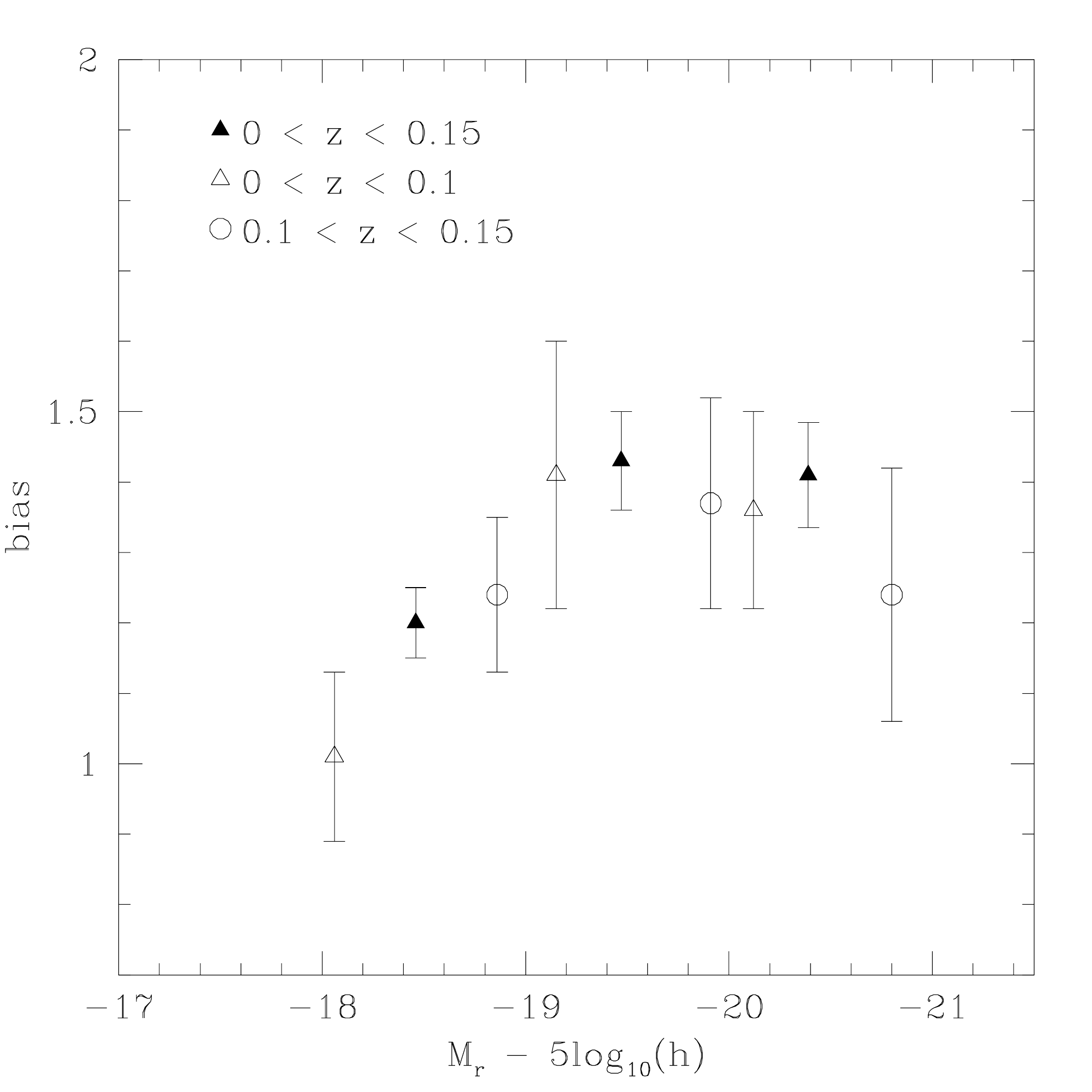}
\caption{The measured bias values using the combination of the photometric auto-correlation and the cross-correlations with the $z_{spec} < 0.15$, $M_r < -20.75$ sample (solid triangles), the cross-correlation with the $z_{spec} < 0.1$, $M_r < -19.75$ (``near"; open triangles), and the cross-correlation with the $0.1 < z_{spec} < 0.15$, $M_r < -20.75$ data (``far"; open circles).}
\label{fig:photbias}
\end{figure}

We display the best-fit $b$ against $M_r$ for galaxies in our photometric redshift samples in Fig.~\ref{fig:photbias}. Values determined from the near cross-correlation are shown by the open triangles, while values determined from the far cross-correlation are shown by the open circles.  The $b$ values determined from the joint fit to the photometric auto-correlations and cross-correlations with the full spectroscopic sample are plotted with solid triangles. Note that the joint fit estimates of $b$ serve as our primary results, while the additional cross-correlations serve as a (correlated) consistency check.

The fact that the results from all cross-correlations agree is important confirmation that photometric redshift distributions we have estimated are accurate.  We note that if we do not include the photometric redshift bias of +0.005, the cross-correlations do not yield consistent results.  For every sample, the cross-correlation with the near data would imply a much lower (by $\sim$ 30\%) value of $b$ and the cross-correlation with the far sample would yield a much higher (by $\sim$ 25\%) value of $b$, implying (the obviously incorrect result) that $b(M_r)$ would be highly oscillatory.  However, this photometric redshift bias does not have a strong effect on the best-fit $b$ values of the three main samples.  When the photometric redshift bias is set to zero, the best-fit $b$ values decrease slightly to $1.19 \pm 0.05$, $1.39 \pm 0.07$, and $1.38 \pm 0.07$ with increasing luminosity.  

We have tested our results further by making changes both to the data we use and the modelling.  We find that our results do not change (within 1$\sigma$) regardless of the minimum angular scale we use to find the best-fit data.  We have investigated multiplying the error on the photometric redshifts by a constant factor and leaving this parameter free when finding the best-fit model.  We find that this can slightly improve the $\chi^2$ values, but not to the extent demanded by removing a degree of freedom.  Importantly, doing so does not change the trend we find with $M_r$ --- the bias of the lowest luminosity red galaxies is always lowest.  We find the same behaviour with different colour cuts and also if we make changes to the redshift/magnitude limits.  No matter the analysis we apply, we find the same result --- red galaxies with $M_r > -19$ have lower bias than their brighter counterparts.  The bias measurements of the photometric galaxies thus suggest that the bias of red galaxies is approximately constant between $-19 > M_r > 21$ and decreases to lower luminosities.  

\section{Bias From Spectroscopic Data} \label{sec:bspec}

\begin{figure*}
\begin{minipage}{7in}
\centering
\includegraphics[width=180mm]{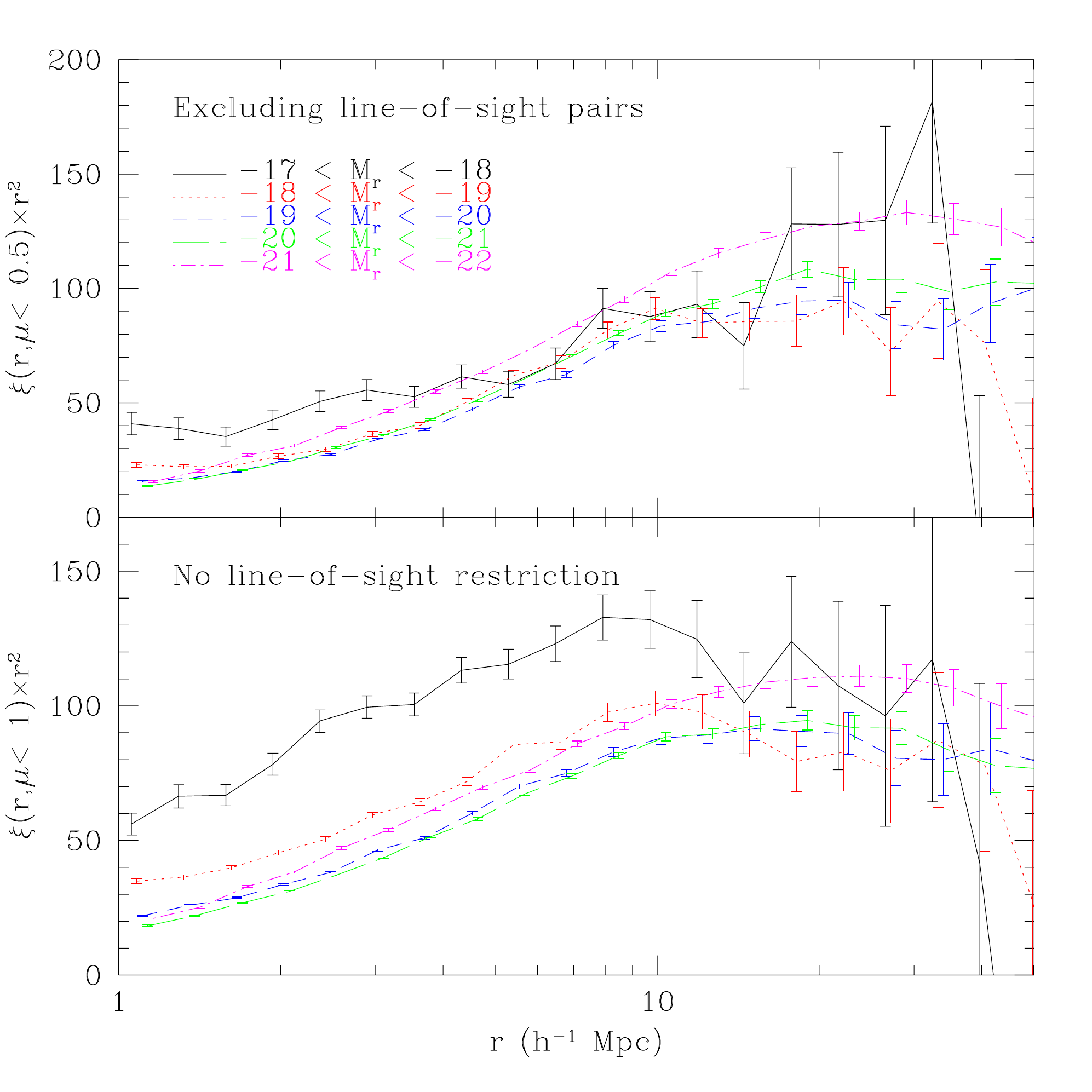}
\caption{The measured redshift-space correlation function multiplied by $r^2$ for the five spectroscopic samples: $-17 < M_r < -18$ (black), $-18 < M_r -19$ (red), $-19 < M_r -20$ (blue), $-20 < M_r < -21$ (green), and $-21 < M_r < -22$ (magenta).  The top panel displays the measurements when we restrict $\mu < 0.5$ and the bottom panel displays the results with no such restrictions.}
\label{fig:ximeas}
\end{minipage}
\end{figure*}

We measure $\xi^s(\mu<\mu_{max},r)$ in order to determine best-fit values of $b$ for the five different spectroscopic samples of red galaxies with magnitudes between $-17$ and $-22$.  The results of these measurements, for $\mu_{max} = 1$ (bottom) and for $\mu_{max} = 0.5$ (top) are displayed in Fig.~\ref{fig:ximeas}.  The two most luminous samples ($-21 < M_r < -22$, magenta; $-20 < M_r < -21$, green) display smooth shapes over the range of the plot and are consistent with respect to each other for both $\mu_{max}$ values.  The least-luminous bin ($-17 < M_r < -18$) is quite noisy and displays a dramatic difference when $\mu_{max}$ is changed --- its clustering amplitudes are roughly double those of the most luminous sample for $r < 10h^{-1}$Mpc when $\mu_{max} = 1$ but this behaviour is removed for $\mu_{max} = 0.5$.  The two samples $-18 < M_r -19$ (red) and $-19 < M_r < -20$ also display significant dependence on the scale and $\mu_{max}$, as both display (relatively) larger amplitudes at smaller scales and greater $\mu_{max}$ values.  We note, however, that any cut that is more restrictive than $\mu < 0.5$ does not produce a statistically significant change in any of our bias measurements.

\begin{figure}
\includegraphics[width=0.9\columnwidth]{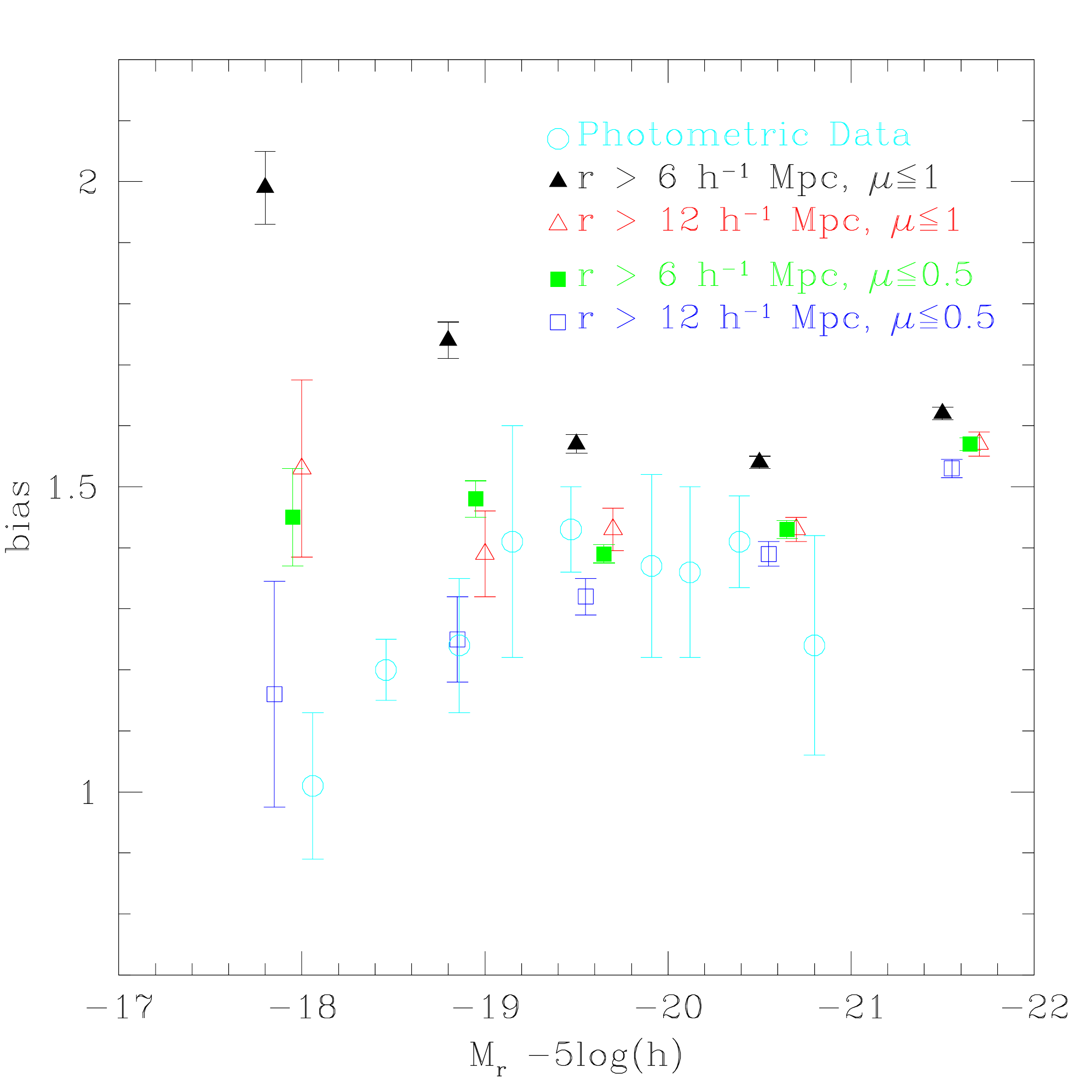}
\caption{The measured bias of red spectroscopic galaxies as a function of $M_r$, for four different choices of $r_{min}$ and $\mu_{max}$.  The $M_r$ values have each been shifted slightly for clarity.  For comparison, the bias values of the red photometric galaxies (same as plotted in Fig.~\ref{fig:photbias}) are displayed with cyan open circles.}
\label{fig:xib}
\end{figure}

Not surprisingly, we find there is a strong dependence both on the minimum $r$ and maximum $\mu$ that we use when we estimate $b$ for the spectroscopic samples.  Fig.~\ref{fig:xib} plots $b$ versus $M_r$ for four separate $r_{min}/\mu_{max}$ limits.  For all bias estimates, the maximum scale we fit to is 40$h^{-1}$Mpc.  If we use all angles to the line of sight and fit above 6$h^{-1}$Mpc (black triangles), we find that the bias increases as the luminosity of the red galaxies decreases.  This behaviour is similar to that reported previously by \cite{swanson} and also \cite{cresswell}.  However, if we fit for $r_{min} > 12h^{-1}$Mpc (open red triangles) or $\mu < 0.5$ (green squares), we see only weak evidence of an increase in the bias towards lower luminosity.  If we require both that $r_{min} > 12h^{-1}$ and $\mu < 0.5$ (open blue squares), we recover a monotonic increase in the bias with luminosity.  The best-fit bias values for $r > 12 h^{-1}$Mpc and $\mu \leq 0.5$ are consistent with the trend we found in the photometric data (plotted in Fig.~\ref{fig:xib} with open cyan circles).

The bias results that we expect to have the least systematic effects are those calculated from measurements at the largest scales (since we expect linear theory to be most valid at larger physical scales) and those calculated for pairs furthest from the line of sight, since they should be the least dependent on redshift space distortions (RSD).  Note that we minimize the effects of linear RSD.  We do allow for linear RSD in our modelling, but studies (see, e.g., \citealt{PW09,Jennings}) have shown that linear RSD models begin to fail at larger scales than real-space clustering models. Perhaps more importantly, we minimize any of Fingers-of-God (FoG) effects \citep{Jackson72}. The effects of FoG imply that, by restricting our analysis to transverse pairs, we are preferentially discounting pairs of galaxies within dark-matter halos.  Note that for the purposes of measuring the linear bias, this is entirely appropriate, as the clustering of galaxies within dark matter halos is clearly non-linear.

We thus believe that, of the data included in Fig.~\ref{fig:xib}, the $r > 12h^{-1}$Mpc, $\mu \leq 0.5$ results are the most trustworthy.  The $\chi^2$/DOF values are all much smaller for the $r > 12h^{-1}$Mpc fits, as for $\mu \leq 1$, the $\chi^2$/DOF values are all greater (and as high as 6) than 2.4 for $r > 6h^{-1}$Mpc and they are all less than 1 for $r > 12h^{-1}$Mpc.  For $\mu \leq 0.5$, the difference is not as extreme, but in every case the $\chi^2$/DOF values are reduced when the fit is performed at larger scales.  For the faintest bin, the bias we measure is sensitive to the particular $\mu_{max}$, $r_{min}$ we choose (it continues to vary significantly if we increase $r_{min}$ beyond $12h^{-1}$Mpc), suggesting that the systematic error associated with this bias measurement is much larger than the statistical error displayed in Fig.~\ref{fig:xib} and subsequent Figures.  

\section{Comparison to Previous Results} \label{sec:com}

\begin{figure}
\includegraphics[width=0.9\columnwidth]{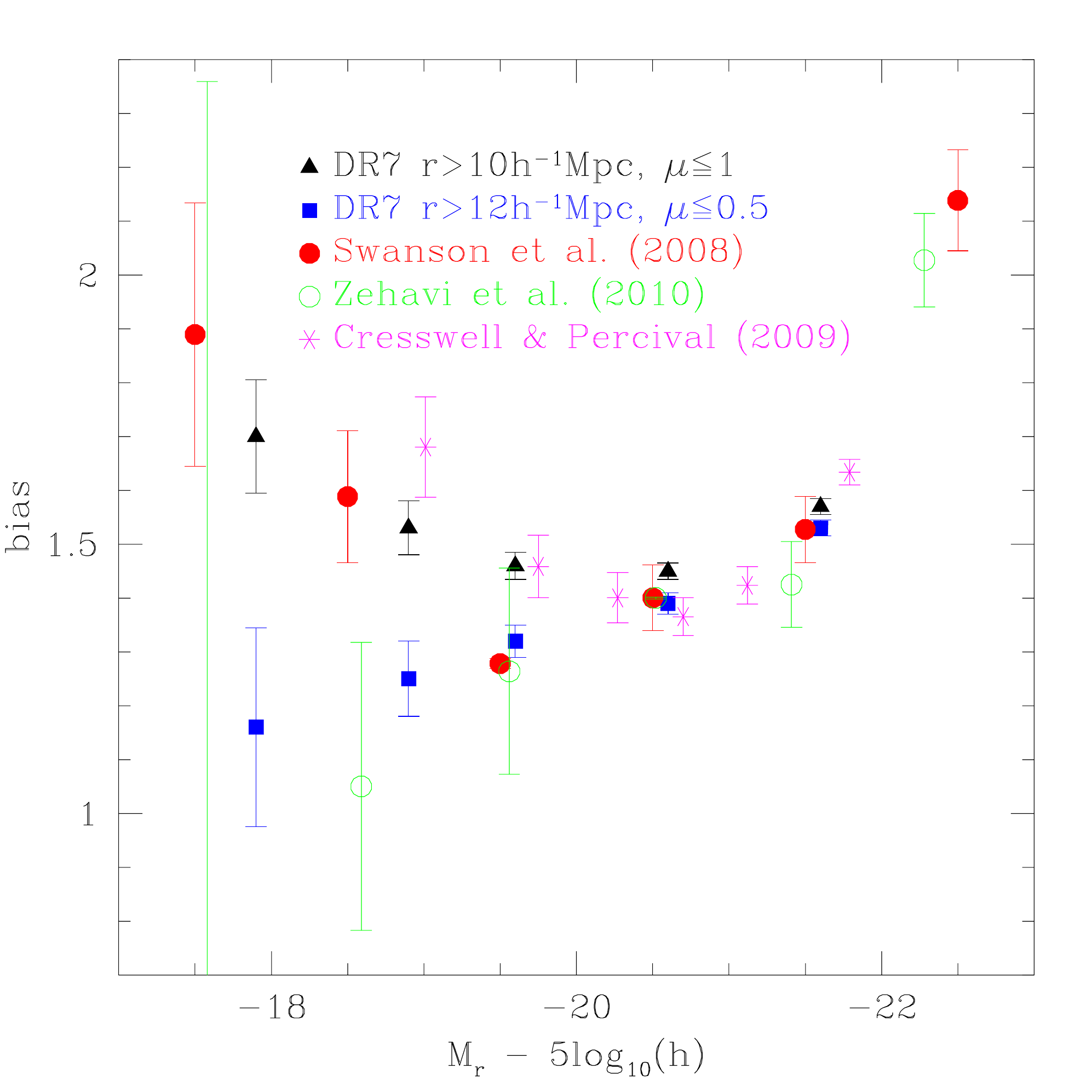}
\caption{The measured bias of red spectroscopic galaxies as a function of $M_r$, for $r > 10 h^{-1}$Mpc, $\mu_{max} = 1$ and $r > 12 h^{-1}$Mpc, $\mu_{max} = 0.5$.  Also plotted are the recent results of three other recent studies of the clustering of red galaxies.}
\label{fig:speccom}
\end{figure}

Our most robust results suggest there is monotonic {\it increase} in the bias of red galaxies with luminosity between $-17.5 > M_r > -19.5$.  This disagrees with recent findings.  A comparison between a selection of our results for the spectroscopic data compared to some previously published results is displayed in Fig.~\ref{fig:speccom}.  We have normalized the bias values of previous studies such that the bias of red galaxies with magnitudes closest to $M_r = -20.5$ is equal to 1.4.  For the \cite{Z10} results, we have used the $w_p(r_p)$ measurements they list in Table B9 at scales $6 < r_p < 17h^{-1}$ Mpc.  We simply determine the inverse-variance weighted mean of the ratio between each of their samples an the $-20 > M_r > -21$ sample and then take its square root to estimate the relative bias of each of their samples.

All of the results plotted in Fig.~\ref{fig:speccom} agree quite well for the most luminous red galaxies with $M_r < -20$. In fact, for $M_r < -19$, the agreement between our spectroscopic analysis with $r_{min} =12h^{-1}$Mpc, $\mu_{max} = 0.5$ (blue squares), \cite{swanson}, and \cite{Z10} is excellent. For less luminous red galaxies, the bias is strongly dependent on the sample and method used.  We find that a choice of $r_{min} =10h^{-1}$Mpc, $\mu_{max} = 1$ (black triangles) for the spectroscopic data analysis produces a result that is quite similar to \cite{cresswell} and \cite{swanson}, while the fit with $r_{min} =12h^{-1}$Mpc, $\mu_{max} = 0.5$ (blue squares) shows strong disagreement.

The dramatic change in the best-fit bias as a function of $r_{min}$ and $\mu_{max}$ suggests that non-linear effects have a great influence on the clustering of faint red galaxies, even to scales $\sim 10h^{-1}$Mpc. The correlation function measurements we perform on the spectroscopic samples (see Fig.~\ref{fig:ximeas}) are consistent with the results of \cite{Hogg03} and \cite{Blanton05}, which found bright and faint red galaxies to have the highest over-densities on 8$h^{-1}$Mpc scales.  \cite{cresswell} fit their power spectra measurements to similarly small scales, as they used $k < 0.21 h$Mpc$^{-1}$. Further, the measurements of \cite{N02} were restricted to $r < 5 h^{-1}$Mpc, and their bias estimates are therefore quite different than ours. Larger scales appear to be required in order to escape the influence of non-linearities.  \cite{swanson} use a larger physical scale (20$h^{-1}$ Mpc) for the radius of their cells, but being that they use counts-in-cells, the counts in large cells are highly covariant with those of smaller scales, and non-linearities may persist.  Further, while all of the studies we cite do attempt to account for redshift-space distortions, none explicitly remove pairs oriented along the line-of-sight, suggesting the possibility that systematic effects may still be present.   

\begin{figure}
\includegraphics[width=0.9\columnwidth]{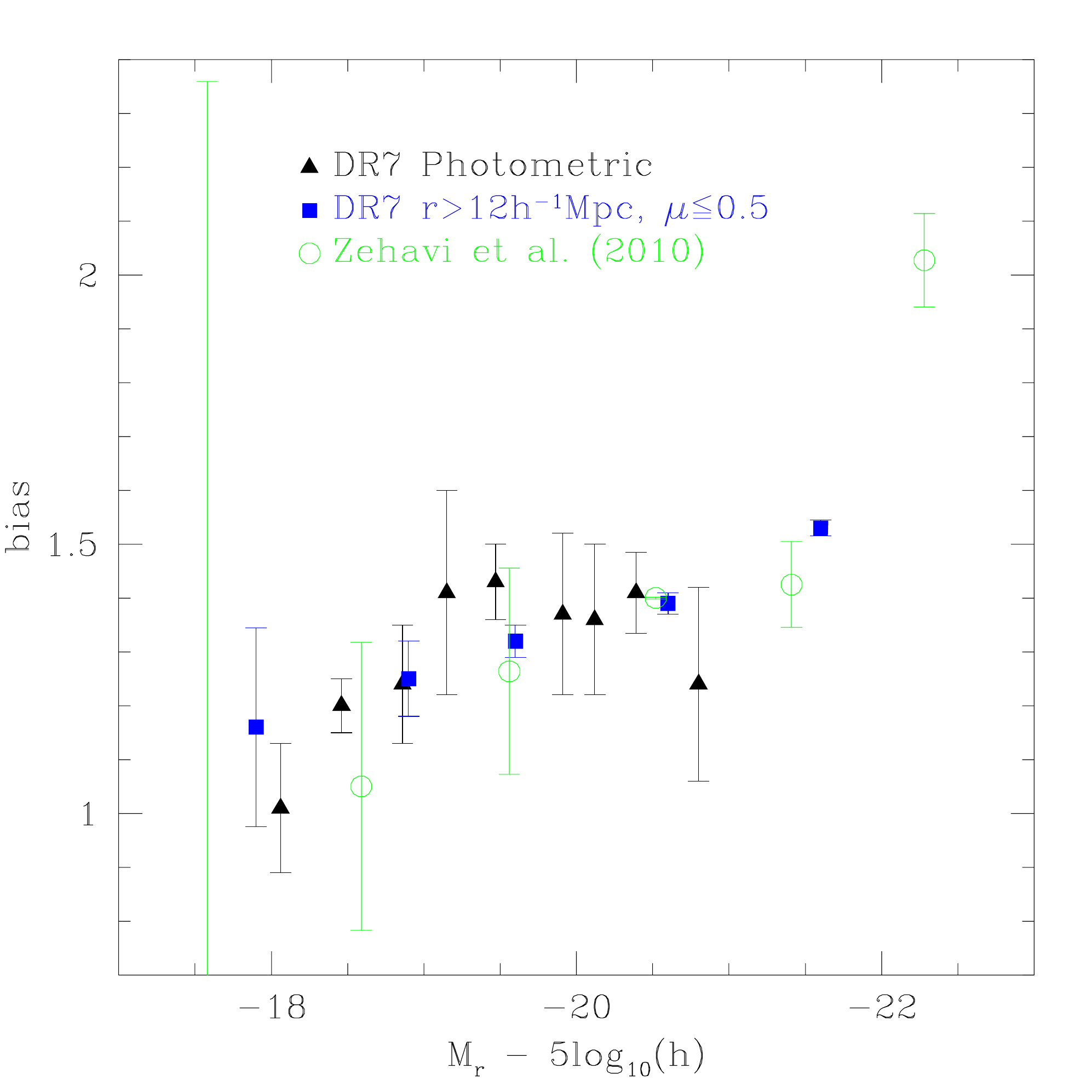}
\caption{The measured bias of red galaxies, considering only DR7 measurements with minimal contribution from pairs along the line of sight.}
\label{fig:DR7bias}
\end{figure}

If we focus only on DR7 data with minimal contribution from clustering along the line of sight, the result is shown in Fig.~\ref{fig:DR7bias} (the $w_p(r_p)$ and $w(\theta)$ measurements should be dominated by transverse pairs).  Within the 1$\sigma$ error-bars, only the measured bias of the $M_r \sim -19.5$ photometric sample appears inconsistent with the rest of the data, but it is consistent to within 2$\sigma$.  Given the number of data points (19), this is reasonable statistically.  The consistency of the data can be further confirmed by the fact that a fit to $b(M_r) = \left(a(M_r+20)\right)^3 + b_o$ has minimum $\chi^2 = 7.2$ for $a=-0.35$ and $b_o = 1.36$ (this suggests only that this data is consistent with a smoothly varying function and do not believe it should be extrapolated outside of the $M_r$ values we present).  The combination of data suggests that the bias of red galaxies increases dramatically for galaxies more luminous than -21.5,  the bias of red galaxies is nearly constant for $-19 > M_r > -21$, but for $M_r > -19$ the bias decreases significantly.

 A reasonable interpretation of the compiled results is that the previously reported increase in the bias of red galaxies as the luminosity decreases is correct only for non-linear scales and therefore not directly related to the average mass of the halos that faint red galaxies occupy.  Non-linear effects were made stronger due to redshift distortions and non-linear clustering, further exacerbated by the very small volume occupied by the low luminosity galaxies. The studies that utilize larger volumes (the photometric data) or use results only where the influence of clustering along the line of sight is minimized (the spectroscopic data herein with $\mu < 0.5$, \citealt{Li,Z10}) agree that the bias of red galaxies displays a monotonic increase with luminosity.

\section{Physical Interpretation} \label{sec:phys}
Our results suggest that redshift distortions and non-linear clustering have a large effect on the measured clustering of faint red galaxies, even at large scales.  Physically, this makes sense.  We know that a significant percentage of faint red galaxies are satellites in high mass halos.  This implies that a significant proportion of the pairs of faint red galaxies are satellites within the same dark matter halo and their true physical separations are as large as a few $h^{-1}$Mpc (the Virial radius of a $10^{14}M_{\odot}/h$ halo is $\sim$ 1 $h^{-1}$Mpc).  Thus, even given the true 3D distribution of galaxies, we would expect to find non-linear effects have more influence on the clustering of faint red galaxies than for other galaxy sub-populations.  

Finger-of-God effects exacerbate the the non-linear effects. Consider a pair of satellite galaxies oriented perpendicularly to the line of sight with a separation of 2$h^{-1}$Mpc:  Assuming they orbit the centre of mass of their dark matter halo at 500km/s, we would interpret their physical separation to be greater than $10h^{-1}$Mpc.  Thus, in redshift space, FoG effects project non-linear clustering to larger scales, and the magnitude of this effect should be largest for faint red galaxies. By ignoring pairs close to the line-of-sight, FoG effects mean that satellite pairs with apparent large-separations are preferentially removed. However, these pairs do not truly have large-separations, and should not contribute to the true large-scale clustering. This explains why removing these pairs brings the results closer in amplitude to the projected clustering results. Given that studies such as \cite{Li} have found that the pairwise-velocity-dispersion is greatest for red galaxies with $M_r < -19$, the FoG effect has its strongest impact on the clustering we measure for the faintest galaxies.  

Previous studies have found an upturn in the large scale bias of red galaxies towards faint luminosities, and this behaviour has been attributed to the fact that these galaxies are predominantly satellites in larger mass haloes (\citealt{Blanton05,swanson, cresswell}). In Section~\ref{sec:intro} we partially motivated this work by pointing out that the implications of \cite{Mas08} and \cite{Wang09} suggested a possible tension with the previous clustering studies. The results we present in this paper show that the reports of an upturn in the bias relationship likely cannot be interpreted as an upturn in the linear bias, due to systematic effects. We find that the large-scale, linear, bias of red galaxies monotonically decreases with decreasing luminosity. This in turn suggests that satellite faint red galaxies, residing in massive haloes, are not dominant by number with respect to central faint red galaxies or faint red satellites of $\sim L*$ galaxies.

Measurements of the satellite fraction in the red cloud vary. \cite{T10} suggest a value of 30\% which does not vary significantly between $z=0.4$ and $z=2$.  \cite{Brown} find that the fraction of stellar mass in satellites increases with host dark matter halo mass, and such that in haloes with $M_h > 10^{14} M_\odot /h$ there is more mass in satellites than centrals. At z=0, they conclude that 30\% of the stellar mass in the red sequence (not number of galaxies) is in satellites. \cite{Wang09}, using a group catalogue and studying very faint (dwarf) red galaxies, conclude that at only around 45\% are satellites. 

 \cite{Z10} suggest an incredibly large satellite fraction of 90\% at the faintest end ($-19 < M_r < -20$), up from 33\% at the intermediate luminosity range ($-21 < M_r < -20$). Interestingly, the HOD model that gives these values (driven by the small-scale clustering) over-predicts the large-scale clustering amplitude. An alternative HOD model, that only allows red satellites in haloes with $M_h > 10^{13} M_\odot /h$, reduces the total satellite fraction in the red cloud to 34\% and provides a significantly better fit to the data at large scales.  Note the apparent contradiction to previous studies; forcing satellite red galaxies to only occupy high mass halos {\it reduces} the large scale bias, due to the fact that such a model allows a lower satellite fraction.  Combined, the results of previous studies suggest that the satellite fraction of red cloud galaxies is $~35\%$ and may increase towards the faint end.  This suggests that, while the satellite fraction allows the bias of red galaxies to remain high relative to blue galaxies, the propensity of faint red galaxies to be central galaxies in low mass halos allows their bias to decrease monotonically.

Central faint red galaxies present an interesting puzzle in galaxy formation. It has been observationally established that the satellite population is in general redder that central galaxies of the same mass (see, e.g., \citealt{vdb08,Yang08,Guo10}). This can be explained by environmental effects as the galaxy falls onto a more massive halo:  a slow or a sudden stripping of the gas reservoir, due to infall, ram-pressure or tidal heating (see, e.g,. \citealt{Gunn72,Larson80,Moore96,Balogh00}). However, a central galaxy has not experienced a direct infall onto a larger halo. One possibility is that put forward by \cite{Wang09}, in which galaxies may experience some form of these environmental effects when passing by a dark matter halo (within 3 times the virial radius), without falling in. Another possibility is simply that the process that shut down star formation in these galaxies in different, or driven by feedback effects. 

\section{Summary and Conclusions} \label{sec:con}

We have computed the large scale bias of red galaxies in the SDSS DR7 as a function of luminosity.  We use both photometric and spectroscopic data to obtain robust results and utilize the largest volume possible.  Our major findings are:

\noindent$\bullet$  The bias of red galaxies we measure in our photometric data implies the faintest red galaxies have the lowest bias (see Fig. \ref{fig:photbias}).  This result is independent of any treatment we apply to the data or its analysis.

\noindent$\bullet$ The bias increases dramatically with increasing luminosity for galaxies with $M_r<-21$. There is no evidence for evolution in the bias for galaxies with $-19>M_r>-21$, and there is weak evidence for a decrease in bias with decreasing luminosity for fainter galaxies (see Figs. 5, 6, and 7).

\noindent$\bullet$  The bias of red galaxies measured in our spectroscopic samples depends greatly on how we choose to analyze the data (see Figs. 5 and 6).  The treatment that we believe is most free of systematics ($r_{min} > 12 h^{-1}$Mpc, $\mu_{max} < 0.5$) recovers a monotonic increase in the bias of red galaxies as a function of luminosity.  This suggests that previous reports of an upturn at low luminosities in the large-scale bias red galaxies were systematically affected by a combination of redshift-space distortions and non-linearities.

The results that we show here, based on the large-scale clustering, give only an incomplete picture of what is happening at the faint end as we cannot disentangle the contribution from central and satellite galaxies to the overall observed bias value. In some sense the bias of faint red galaxies represents the interplay between the number of satellites (with strong bias, in high mass haloes), and centrals (with low bias, in low mass haloes). Nonetheless our results are in good qualitative agreement with other measurements of the fraction of red satellites in the sense that the fraction of satellites is not expected to be dominant even at the faint end. Treatment of the clustering of red galaxies utilizing halo-occupation-distribution modelling to constrain the evolution and mass-assembly of red galaxies will be presented in a follow-up paper.

\section{Acknowledgments}
We acknowledge Molly Swanson, Zheng Zheng, and Peder Norberg for helpful comments and discussions.  We acknowledge an anonymous referee for useful comments.

AJR and WJP thank the UK Science and Technology Facilities Council for financial support; RT and WJP are also grateful to the Leverhulme trust and the European Research Council. 

Funding for the creation and distribution of the SDSS Archive has been provided by the Alfred P. Sloan Foundation, the Participating Institutions, the National Aeronautics and Space Administration, the National Science Foundation, the U.S. Department of Energy, the Japanese Monbukagakusho, and the Max Planck Society. The SDSS Web site is http://www.sdss.org/.

The SDSS is managed by the Astrophysical Research Consortium (ARC) for the Participating Institutions. The Participating Institutions are the University of Chicago, Fermilab, the Institute for Advanced Study, the Japan ParticipationGroup, Johns Hopkins University, the Korean Scientist Group, Los Alamos National Laboratory, the Max Planck Institute for Astronomy (MPIA), the Max Planck Institute for Astrophysics (MPA), New Mexico State University, the University of Pittsburgh, the University of Portsmouth, Princeton University, the United States Naval Observatory, and the University of Washington.


\begin{thebibliography}{7}

\bibitem[\protect\citeauthoryear{Abazajian et al.}{2009}]{DR7} 
  Abazajian, K.~N., et al., 2009, ApJS, 182, 543

\bibitem[\protect\citeauthoryear{Ball et al.}{2008}]{Ball08} 
  Ball, N.~M., Loveday, J., Brunner, R.~J., 2008, MNRAS, 383, 907

\bibitem[\protect\citeauthoryear{Balogh \& Morris}{2000}]{Balogh00} 
  Balogh, M.~L., Morris, S.~L., 2000, MNRAS, 318, 703

\bibitem[\protect\citeauthoryear{Bardeen et al.}{1986}]{BBKS}
  Bardeen J.M., Bond J.R., Kaiser N., Szalay A.S., 1986, ApJ, 
  304, 15

\bibitem[\protect\citeauthoryear{Bell et al.}{2004}]{Bell04} 
  Bell, E.~F., et al., 2004, ApJ, 608, 752

\bibitem[\protect\citeauthoryear{Berlind et al.}{2005}]{Berlind05} 
  Berlind, A.~A., Blanton, M.~R., Hogg, D.~W., Weinberg, D.~H., 
  Dav{\'e}, R., Eisenstein, D.~J., Katz, N., 2005, ApJ, 629, 625

\bibitem[\protect\citeauthoryear{Blanton et al.}{2003}]{BlantonEtAl03} 
  Blanton, M.~R., et al., 2003, ApJ, 594, 186

\bibitem[\protect\citeauthoryear{Blanton et al.}{2005}]{Blanton05} 
  Blanton, M.~R., Eisenstein, D., Hogg, D.~W., Schlegel, D.~J., 
  Brinkmann, J., 2005, ApJ, 629, 143

\bibitem[\protect\citeauthoryear{Brown et al.}{2008}]{Brown} 
  Brown, M. J. I. et al., 2008, ApJ, 682, 937 

\bibitem[\protect\citeauthoryear{Budav{\'a}ri et al.}{2003}]{Bud03} 
  Budav{\'a}ri, T., et al., 2003, ApJ, 595, 59

\bibitem[\protect\citeauthoryear{Coil et al.}{2006}]{Coil06} 
  Coil, A.~L., Newman, J.~A., Cooper, M.~C., Davis, M., 
  Faber, S.~M., Koo, D.~C., Willmer, C.~N.~A., 2006, ApJ, 644, 671
 
\bibitem[\protect\citeauthoryear{Cole \& Kaiser}{1989}]{cole89} 
  Cole S., Kaiser N., 1989, MNRAS, 237, 1127

\bibitem[\protect\citeauthoryear{Conroy et al.}{2007}]{ConroyEtAl07} 
  Conroy, C., Ho, S., White, M., 2007, MNRAS, 379, 1491

\bibitem[\protect\citeauthoryear{Cooray \& Sheth}{2002}]{CooSh02} 
  Cooray, A., Sheth, R., 2002, Phys. Rep., 372, 1

\bibitem[\protect\citeauthoryear{Cowie et al.}{1996}]{Cow96} 
  Cowie, L.~L., Songaila, A., Hu, E.~M., Cohen, J.~G., 1996, AJ, 112, 839

\bibitem[\protect\citeauthoryear{Cresswell \& Percival}{2009}]{cresswell} 
  Cresswell, J.~G., Percival, W.~J., 2009, MNRAS, 392, 682

\bibitem[\protect\citeauthoryear{Croton et al.}{2006}]{Cr06} 
  Croton, D.~J., Norberg, P., Gazta{\~n}aga, E.,
  Baugh, C.~M., 2007, MNRAS, 379, 1562

\bibitem[\protect\citeauthoryear{Dressler}{1980}]{Dre80} 
  Dressler,~A., 1980, ApJ, 236, 351

\bibitem[\protect\citeauthoryear{Eisenstein \& Hu}{1998}]{EH} 
  Eisenstein, D.~J., Hu, W., 1998, ApJ, 496, 605

\bibitem[\protect\citeauthoryear{Feldman et al.}{1994}]{fkp} 
  Feldman, H.~A., Kaiser, N., Peacock, J.~A., 1994, ApJ, 426, 23

\bibitem[\protect\citeauthoryear{Fukugita et al.}{1996}]{F} 
  Fukugita, M., Ichikawa, T., Gunn, J.~E., Doi, M., Shimasaku, K.,  
  Schneider, D.~P., 1996, AJ, 111, 1748

\bibitem[\protect\citeauthoryear{G{\'o}rski et~al.}{2005}]{GorskiEtAl05}
  G{\'o}rski, K.~M., Hivon, E., Banday, A.~J., Wandelt, B.~D., 
  Hansen, F.~K., Reinecke, M., Bartelmann, M., 2005, ApJ, 622, 759
  
\bibitem[\protect\citeauthoryear{Gunn \& Gott}{1972}]{Gunn72} 
  Gunn, J.~E., Gott, J.~R., III, 1972, ApJ, 176, 1
  
\bibitem[\protect\citeauthoryear{Gunn et al.}{1998}]{C} 
  Gunn, J.~E., et al., 1998, AJ, 116, 3040
  
\bibitem[\protect\citeauthoryear{Guo et al.}{2010}]{Guo10} 
  Guo, Q., White, S., Li, C., Boylan-Kolchin, M., 2010, MNRAS, 404, 1111

\bibitem[\protect\citeauthoryear{Haines et al.}{2007}]{Haines} Haines, C.~P., Gargiulo, 
A., La Barbera, F., Mercurio, A., Merluzzi, P., \& Busarello, G.\ 2007, MNRAS, 381, 7

\bibitem[\protect\citeauthoryear{Hamilton}{1992}]{hamilton92} 
  Hamilton, A.J.S., 1992, ApJ, 385, L5

\bibitem[\protect\citeauthoryear{Hogg et al.}{2003}]{Hogg03} 
  Hogg, D.~W., et al., 2003, ApJL, 585, L5

\bibitem[\protect\citeauthoryear{Jackson}{1972}]{Jackson72} 
    Jackson, J.~P.\ 1972, Ph.D.~Thesis

\bibitem[\protect\citeauthoryear{Jennings et al.}{2010}]{Jennings} Jennings, E., Baugh, 
C.~M., \& Pascoli, S.\ 2010, MNRAS 1572

\bibitem[\protect\citeauthoryear{Kauffmann et al.}{1997}]{Kauf97} 
  Kauffmann, G., Nusser, A., Steinmetz, M., 1997, MNRAS, 286, 795

\bibitem[\protect\citeauthoryear{Landy \& Szalay}{1993}]{LZ} 
  Landy, S. D., Szalay, A. S., 1993, ApJ, 412, 64

\bibitem[\protect\citeauthoryear{Larson et al.}{1980}]{Larson80} 
  Larson, R.~B., Tinsley, B.~M., Caldwell, C.~N., 1980, ApJ, 237, 692

\bibitem[\protect\citeauthoryear{Li et al.}{2006}]{Li} 
  Li, C., Kauffmann, G., Jing, Y. P., White, S. D. M., 
  Borner, G., Cheng, F. Z., 2006, MNRAS, 368, 21

\bibitem[\protect\citeauthoryear{Madgwick et al.}{2003}]{Ma03} 
  Madgwick, D.~S., et al., 2003, MNRAS, 344, 847 

\bibitem[\protect\citeauthoryear{Masjedi et al.}{2008}]{Mas08} 
  Masjedi, M., Hogg, D.~W., Blanton, M.~R., 2008, ApJ, 679, 260

\bibitem[\protect\citeauthoryear{McCracken et al.}{2007}]{Mc07} 
  McCracken, H.~J., et al., 2007, ApJS, 172, 314

\bibitem[\protect\citeauthoryear{McCracken et al.}{2008}]{M08} 
  McCracken, H.~J., Ilbert, O., Mellier, Y., Bertin, E., 
  Guzzo, L., Arnouts, S., Le F{\`e}vre, O., Zamorani, G.,
  2008, A\&A, 479, 321

\bibitem[\protect\citeauthoryear{Moore et al.}{1996}]{Moore96} 
  Moore, B., Katz, N., Lake, G., Dressler, A., Oemler, A.,
  1996, Nature, 379, 613

\bibitem[\protect\citeauthoryear{Myers et al.}{2007}]{Mye07} 
  Myers, A.~D., et al., 2007, ApJ, 658, 85 

\bibitem[\protect\citeauthoryear{Norberg et al.}{2002}]{N02} 
  Norberg, P., et al., 2002, MNRAS, 332, 827

\bibitem[\protect\citeauthoryear{Peacock \& Smith}{2000}]{PS00} 
  Peacock, J.~A., Smith R.~E., 2000, MNRAS, 318, 1144

\bibitem[\protect\citeauthoryear{Percival \& White}{2009}]{PW09} Percival, W.~J., \& White, M.\ 2009, MNRAS, 393, 297

\bibitem[\protect\citeauthoryear{Petrosian}{1976}]{Petrosian} 
  Petrosian, V., 1976, ApJ, 210, L53

\bibitem[\protect\citeauthoryear{Reid et al.}{2009}]{reidetal09} 
  Reid, B. A., et al., 2009, MNRAS, 404, 60

\bibitem[\protect\citeauthoryear{Ross et al.}{2007}]{R07} 
  Ross, A.~J., Brunner, R.~J., Myers, A.~D., 2007, ApJ, 665, 67 (R07)

\bibitem[\protect\citeauthoryear{Ross et al.}{2009}]{R09} 
  Ross, A.~J., Brunner, R.~J., 2009, MNRAS, 399, 878

\bibitem[\protect\citeauthoryear{Ross et al.}{2010}]{R10} 
  Ross, A.~J., Percival W.~J., Brunner, R.~J., 
  2010, MNRAS, 407, 420 (R10)

\bibitem[\protect\citeauthoryear{Scranton et al.}{2002}]{Scr02} 
  Scranton, R., et al., 2002, ApJ, 579, 48

\bibitem[\protect\citeauthoryear{Skibba et al.}{2009}]{Skibba08} 
  Skibba, R.~A., et al., 2009, MNRAS, 399, 966

\bibitem[\protect\citeauthoryear{Smith et al.}{2003}]{smith} 
  Smith, R.~E., et al., 2003, MNRAS, 341, 1311

\bibitem[\protect\citeauthoryear{Swanson et al.}{2008}]{swanson} 
  Swanson, M.~E.~C., Tegmark, M., Blanton, M., Zehavi, I., 2008, 
  MNRAS, 385, 1635

\bibitem[\protect\citeauthoryear{Tegmark}{1997}]{tegfish} 
  Tegmark, M., 1997, Physical Review Letters, 79, 3806

\bibitem[\protect\citeauthoryear{Tinker et al.}{2005}]{Tinker} 
  Tinker J. L., Weinberg D. H., Zheng Z., Zehavi I., 2005, ApJ, 631, 41

\bibitem[\protect\citeauthoryear{Tinker et al.}{2008}]{Tink08} 
  Tinker, J.~L., Conroy, C., Norberg, P., Patiri, S.~G., Weinberg, D.~H., 
  Warren, M.~S., 2008, ApJ, 686, 53

\bibitem[\protect\citeauthoryear{Tinker \& Wetzel}{2010}]{T10} 
  Tinker, J.~L., Wetzel, A.~R., 2010, ApJ, 719, 88

\bibitem[\protect\citeauthoryear{Wake et al.}{2008}]{Wake08} 
  Wake, D.~A., et al., 2008 MNRAS, 387, 1045

\bibitem[\protect\citeauthoryear{Wang et al.}{2009}]{Wang09} 
  Wang, Y., Yang, X., Mo, H.~J., van den Bosch, F.~C., Katz, N., 
  Pasquali, A., McIntosh, D.~H., Weinmann, S.~M., 2009, ApJ, 697, 247

\bibitem[\protect\citeauthoryear{van den Bosch et al.}{2008}]{vdb08} 
  van den Bosch, F.~C., Aquino, D., Yang, X., Mo, H.~J., Pasquali, A., 
  McIntosh, D.~H., Weinmann, S.~M., Kang, X., 2008, MNRAS, 387, 79

\bibitem[\protect\citeauthoryear{Willmer et al.}{1998}]{W98} 
  Willmer, C.~N.~A., da Costa, L.~N., Pellegrini, P.~S., 
  1998, AJ, 115, 869

\bibitem[\protect\citeauthoryear{White \& Rees}{1978}]{WhiteRees1978} 
  White, S.~D.~M., Rees, M.~J., 1978, MNRAS, 183, 341

\bibitem[\protect\citeauthoryear{White et al.}{2007}]{MW07} 
  White, M., Zheng, Z., Brown, M.~J.~I., Dey, A., 
  Jannuzi, B.~T., 2007, ApjL, 655, L69

\bibitem[\protect\citeauthoryear{Yang et al.}{2008}]{Yang08} 
  Yang, X., Mo, H.~J., van den Bosch, F.~C., 2008, ApJ, 676, 248

\bibitem[\protect\citeauthoryear{Zehavi et al.}{2004}]{Z04} 
  Zehavi, I., et al., 2004, ApJ, 608, 16

\bibitem[\protect\citeauthoryear{Zehavi et al.}{2005}]{Z05} 
  Zehavi, I., et al., 2005, ApJ, 630, 1 

\bibitem[\protect\citeauthoryear{Zehavi et al.}{2010}]{Z10} 
  Zehavi I., et al., 2010, arXiv:1005.2413

\bibitem[\protect\citeauthoryear{Zheng et al.}{2005}]{zheng05} 
  Zheng, Z., et al., 2005, ApJ, 633, 791

\bibitem[\protect\citeauthoryear{Zheng et al.}{2007}]{ZZ07} 
  Zheng, Z., Coil, A.~L., Zehavi, I., 2007, ApJ, 667, 760

\label{lastpage}
\end{thebibliography}
\end{document}